\documentclass[10pt,aps,preprintnumbers,prd,noshowpacs,nofootinbib,noshowkeys,floatfix,superscriptaddress]{revtex4-2}
\usepackage[dvips]{graphics,graphicx}
\usepackage[colorlinks=true,linktocpage=true,linkcolor=blue,citecolor=blue]{hyperref}
\usepackage[usenames,dvipsnames]{color}
\usepackage{amsmath, amssymb,oldgerm}
\usepackage{multirow}
\usepackage{longtable}                                                                                                                  \usepackage{color}
\usepackage[normalem]{ulem}  
\usepackage{braket}
\usepackage{slashed}
\usepackage[mathscr]{eucal}

\newcommand{\n}{\nonumber}

\newcommand{\C}{\mathcal{C}}

\newcommand{\beq}{\begin{equation}}
\newcommand{\eeq}{\end{equation}}

\newcommand{\eq}{Eq.~}

\newcommand{\inlangle}{{\phantom{\Big|}}_\text{in}\Big\langle}
\newcommand{\inrangle}{\Big\rangle_\text{in}}
\newcommand{\inlanglesm}{{}_\text{in}\langle}
\newcommand{\inranglesm}{\rangle_\text{in}}
\newcommand{\outlangle}{{\phantom{\Big|}}_\text{out}\Big\langle}
\newcommand{\outrangle}{\Big\rangle_\text{out}}
\newcommand{\outlanglesm}{{}_\text{out}\langle}

\newcommand{\os}{\text{on-shell}}
\newcommand{\osl}{\text{on-shell},l}

\begin{document}

\title{Relativistic kinetic equation for dense gases  from quantum field theory}

\author{Jin Hu}
\email{hu-j17@mails.tsinghua.edu.cn}
\affiliation{Department of Physics, Tsinghua University, Beijing 100084, China}

\begin{abstract}
We derive the relativistic kinetic equation and collision kernel  for dense gases of spin $0$ particles from quantum field theory based on the Wigner-function formalism. The formalism developed by Degroot can be used as an effective way for density expansion of the kinetic equation. 
The kinetic equation obtained in the lowest order in density recovers the Boltzmann equation with a nonlocal binary collision term. Keeping this expansion procedure, we derive the triple collision term, which can be seen as a relativistic  extension of correspondent  works on the  transport research about dense gases in the non-relativistic cases. Considering the widespread practicability in different physical systems, the kinetic equation  for dense gases we obtain shall be put into good use.
\end{abstract}




\maketitle

\section{Introduction}
 The research on dense gases has long attracted the interests of physicists. The first important attempt extending Boltzmann to higher density was made by Enskog \cite{Brush, Chap}, who takes nonlocality of collisions into account by introducing position shifts of particles taking part in the collision  from the collision center. Such a nontrivial extension leads to the transport coefficents in good agreement with  experimental results over a wide range of densities. However, the Enskog theory is confined to the description of hard sphere potential and thus not a complete theory. Historially, Bogoliubov turned the related research work into a systematic theory \cite{Bogoliubov}, which lays  the foundation for the modern kinetic theory. Uhlenbeck further developed Bogoliubov's ideas and introduced the concepts of approximating the dynamics of dense gases via the expansion of small groups of particles. The cooperation with Choh led to the implementation of this original thinking. In Choh's thesis \cite{Choh}, they worked out the transport equation with a triple collision kernel, which makes  the application to dense gases plausible and this pioneering work was well reviewed by  \cite{Cohen,Dorfman,ernst}. Similar kinetic theory and further  discussions and developments  can  be also found in \cite{greena, greenb,greenc} and \cite{Cohena, Cohenb,Cohenc,Cohend,Cohene}. In the 1980's, the authors of  \cite{deboerw, deboerv} developed the related works and obtained a triple collision term appropriate for nonrelativistic quantum statistics.

Considering the above discussion is restricted to nonrelativistic cases, an extension to the relativistic kinetic equation for dense gases is necessary especially for talking about dense  matters created in high energy processes. Such processes include but are not limited to relativistic heavy ion or nucleons collisions, compact stars or galaxies, and the so-called lepton era in the evolution of the early universe. In these processes, density effects play a big role in the transport phenomena of related physical systems and correspondent transport coefficients will receive considerable corrections from density expansion about which the formal introduction can be found in \cite{Dorfman, Dorfmann}. To elucidate the mystery of high density transport, the framework on the kinetic equation for dense gases should be lay importance on, from which one can establish a quantitative numerical model or seek a quallitative undersatnding. Last but not the least, a dense kinetic equation will be helpful for our finding realistic equation of states, for instance, a systematic research on Enskog kinetic equation provides us with the EOS of Van der Waals gases, which is really a huge progress in statistical mechanics and kinetic theory.
In this work, we perform  a detailed derivation of relativistic kinetic equation for dense gases based on the framework developed by Degroot \cite{DeGroot:1980dk}, from which the Boltzmann equation can be constructed using the first-principle formulation. To note, the series of works \cite{deboerw, deboerv}  on the triple collision term have common ground with Degroot's formalism in many aspects.

The paper is organized as follows. In Sec.\ \ref{secqu}, we derive the quantum 
transport equation from the Wigner-function formalism. 
In Secs.\ \ref{secloc} and \ref{secnon} we explicitly derive the local and nonlocal parts of 
the collision term respectively with only binary collisions involved. Such calculations are based on a specific expansion scheme i.e., gradients expansion, and we cut the expansion up to $\mathcal{O}(\hbar)$. The obtained equation recovers the nonlocal form of usual Boltzmann equation with binary collisions. In Sec.\ \ref{dense}, we apply the techniques developed in \cite{DeGroot:1980dk} to derive the triple collision term with the contributions belonging to higher order in density neglected. In this section, we do not seek gradients expansion similar to what we did in Secs.\ \ref{secloc} and \ref{secnon}. Finally, conclusions and outlook are given in Sec.\ \ref{secconc}. 
We use the following notation and conventions: $a\cdot b=a^\mu b_\mu$,
$g_{\mu \nu} = \mathrm{diag}(+,-,-,-)$, and repeated indices are summed over.

\section{Quantum transport equations}
\label{secqu}

In this work, we focus on the form of collision term for dense gases and consider only spinless particles, whcih is free from the complexity of handling spinor calculations. The Wigner function for spin-0 particles is defined as
\cite{DeGroot:1980dk},
\begin{equation}
  W(x,p)=4\pi\int \frac{d^4y}{(2\pi\hbar)^5} e^{-\frac i\hbar p\cdot y}
  \left\langle :\phi^\dagger\left(x_1\right)\phi\left(x_2\right):\right\rangle \; , \label{wigdef}
\end{equation}
with $x_{1,2}=x \pm y/2$ and $\phi (x),\, \phi^\dagger(x)$ being  Klein Gordon  fields. {Here we use $\langle:\, :\rangle$  to denote the normal-ordered ensemble average.}
 Without losing generality, the Lagrangian for Klein Gordon fields  is chosen to be of the form
\begin{equation}
\mathcal{L}=-\frac{\hbar}{2}\left(\phi^\dagger\partial^2\phi+\phi\partial^2\phi^\dagger+\frac{2m^2}{\hbar^2}\phi^\dagger\phi\right)
+\mathcal{L}_I\;, 
\end{equation}
 {If there are interactions coupling to gauge field in $\mathcal{L}_I$, the Wigner function Eq.(\ref{wigdef})  must be modified with a gauge link in the definition \cite{Vasak:1987um}.} 
Derivative of  the Lagrangian with respect to $\phi^\dagger$ gives the equation of motion for $\phi$,
\begin{align}
 \left(\partial^2+\frac{m^2}{\hbar^2}\right)\phi(x)&=\rho(x)\; , \label{Dirac}
\end{align}
where $\rho= (1/ \hbar) \partial \mathcal{L}_I /\partial \phi^\dagger $ in the case of non-derivative coupling. Similar manipulation can be applied to get the equation for the congugate field $\phi^\dagger$.
Starting from Eq.~\eqref{Dirac} we derive 
 a transport equation for the 
Wigner function,
\begin{equation}
p\cdot\partial\, W(x,p)=\C(x,p) \;, \label{wigboltz9}
\end{equation}
with{
\begin{eqnarray}
\label{cwww}
\C&=&i\int \frac{d^4y}{(2\pi\hbar)^4} e^{-\frac i\hbar p\cdot y}
\left\langle :\phi^\dagger(x_1)\rho\left(x_2\right)-\rho^\dagger(x_1)\phi\left(x_2\right) :\right\rangle\; ,
\end{eqnarray}}
and a modified on-shell condition \cite{DeGroot:1980dk}
\begin{eqnarray}
\left(m^2-p^2+\frac{\hbar^2}{4}\partial^2\right)W(x,p)
&=& \hbar\,\delta M(x,p) \; ,
\label{onshellwiggg}
\end{eqnarray}
with
{
	\begin{eqnarray}
	\delta M&\equiv&\int \frac{d^4y}{(2\pi\hbar)^4}\, 
	e^{-\frac{i}{\hbar}p\cdot y}
\left\langle :\rho^\dagger(x_1)\phi\left(x_2\right)+\phi^\dagger(x_1)\rho\left(x_2\right) :\right\rangle\; .
	\label{deltaM}
	\end{eqnarray}}
For simplicity, we  restrict our discussions to the positive-energy part of the
Wigner function and we do not intend to denote this point with a heaviside function $\theta(p_0)$. The extension to negative energies follows the same footing and can be put in a  straightforward fashion.  And we only consider a simple system, i.e., only one particle species and elastic collisions between partilces are involved.

From now on, we apply Degroot's formalism to derive the collision term~\cite{DeGroot:1980dk} based on reduction formulas and operator expansions with asymptotic particle states.  The noninteracting initial $n$-particle states are defined 
as 
\begin{equation}
\label{statein}
|p_1,\ldots,p_n\rangle_\text{in}\equiv a^\dagger_{\text{in}}(p_1)\cdots 
a^\dagger_{\text{in}}(p_n)|0\rangle \; ,
\end{equation}
where $p_i$, $i=1, \ldots, n$, denote the particle momentum  respectively, and 
$a^\dagger_{\text{in}}(p_i)$ is the creation operator for that particle. According to Bogoliubov~\cite{Bogoliubov}, the molecular-chaos assumption should be treated as the initial condition to seek higher density formulation, which characterizes the kinetic stage of nonequilibrium system evolution and reveals vanishing particle correlations after several collision times. The only remaining correlation is statistical correlation, in other words, the exchange symmetry of Bosons. With these considerations, the density expansion can be made order by order. First, the discussions are confined  to two-particle states, 
i.e.,  binary collisions. 
With this simplification, Eq.\ \eqref{cwww} can be shown as the form \cite{DeGroot:1980dk} 
\begin{align}
C_{22}={}& \frac{1}{(2\pi\hbar)^2}
\int d^4 x_{1}d^4 x_{2}d^4 p_{1}d^4 p_{2}d^4 q_{1}d^4 q_{2}
{\phantom{\Big|}}_\text{in}
\Big\langle{p_1-\frac{q_1}{2},p_2-\frac{q_2}{2} \Big| \Phi(p) 
\Big| p_1+\frac{q_1}{2}, p_2+\frac{q_2}{2}}\Big\rangle_\text{in}\n\\
&\times\prod_{j=1}^2 \exp\left(\frac{i}{\hbar}q_j\cdot x_j\right)
  W_\text{in}(x+x_j, p_j)\; ,    \label{Wcollttt}
\end{align}
where the operator $\Phi$ is given by
\begin{align}
\Phi(p)& \equiv \frac{i}{2}\int\frac{d^4y}{(2\pi\hbar)^4} 
e^{-\frac i\hbar p\cdot y} 
:\Big( \phi^\dagger\left(\frac y2\right)  
\rho\left(-\frac y2\right)-
\rho^\dagger\left(\frac y2\right)  
\phi\left(-\frac y2\right)\,\Big):\;,
\label{phigen}
\end{align}
where $P^\mu$ is the total 4-momentum operator. A comment is followed that $C_{22}$ together with Eq.(\ref{ass1}) constitutes the double collision term originated from two particles states and more details are postponed afterwards. 
To close the Boltzmann-type equation \eqref{wigboltz9} for the interacting Wigner function $W$ with the collision kernel 
\eqref{Wcollttt}, we need to reexpress $W_{\text{in}}$ in terms of $W$. This can be achieved for a dilute system in the following way,
\begin{align}
\label{ass1}
W&=W_\text{in}+ \ldots ,
\end{align} 
where the ellipsis represents corrections of higher order in density 
following \cite{DeGroot:1980dk}.
 Replacing $W_\text{in}$ in the collision term with $W$ neglecting the corrections within the ellipsis, the closed transport equation for dilute gases can be persued.
In general,  the collision term in Eq.\ \eqref{Wcollttt} is nonlocal and noninstant, as can be seen from their dependance on $x+x_j$. In order to exhibit the local form of well-known Boltzmann equation, one can Taylor-expand $W(x+x_j, p_j)$ 
around $x$ assuming that the Wigner function varies slowly in space and time on the microscopic scale in view of short-range interaction. There is no doubt that such an implementation leads to the local Boltzmann equation familiar to us shown in  \cite{DeGroot:1980dk} and the review in the next section.  Also, the term of first order in gradients can be kept
(equivalent to first order in $\hbar$), and the authors of \cite{Weickgenannt:2021cuo} take this as a starting point to derive nonlocal collision term for massive spin-$1/2$ particles, i.e.,
\begin{equation}
\label{ass2}
W(x+x_j, p_j)=W(x, p_j)+x_j \cdot \partial W(x,p_j) \;.
\end{equation}
Substitutition of  Eqs.\ \eqref{ass1} and \eqref{ass2} into Eq.\ \eqref{Wcollttt} gives,
\begin{align}
C_{22} ={}& (2\pi\hbar)^6
\int d^4 p_{1}d^4 p_{2} d^4 q_{1}d^4 q_{2} \, {\phantom{\Big|}}_\text{in}
\Big\langle{p_1-\frac{q_1}{2},p_2-\frac{q_2}{2}\Big|\Phi(p)
\Big|p_1+\frac{q_1}{2}, p_2+\frac{q_2}{2}}\Big\rangle_\text{in}\n\\
  &\times\prod_{j=1}^2  
  \left\{W(x, p_j)\delta^{(4)}(q_j)
  -i\hbar\left[\partial_{q_j}^\mu\delta^{(4)}(q_j)\right]\partial_{\mu}W(x,p_j)\right\}
 \; , 
  \label{generalcollisionterm}
 \end{align}
where  the integration over $d^4x_1$ and $d^4 x_2$ is performed.  The explicit form of $\hbar\partial^\mu_{q_j}$ reveals the equilvalence of $\hbar$ expansion and gradients expansion. As will be presented in follow-up, the first and second term in the second line of Eq.(\ref{generalcollisionterm}) represents the local and nonlocal collision term respectively.

\section{Local collisions for dilute gases}
\label{secloc}

This section can be seen as  a review work of correspondent contents in~\cite{DeGroot:1980dk}. The authors discard the nonlocal part in the second line of Eq.(\ref{generalcollisionterm}). We follow their practice to derive local collision term.
The in-matrix element of $\Phi$ within Eq.(\ref{generalcollisionterm}), with $\Phi$ defined in  
Eq.~\eqref{phigen}, is explicitly calculated in App.~\ref{collkerapp}. 
Then the  local part of the collision term is
\begin{align}
&(2\pi\hbar)^6\inlanglesm p_1,p_2|\Phi| p_1,p_2\inranglesm 
=2\delta(p^2-m^2)\Bigg\{ \int dP^\prime 
\delta(p+p^\prime-p_1-p_2) \langle{p,p^\prime|t|p_1,p_2}\rangle
\langle{p_1,p_2|t^\dagger|p,p^\prime}\rangle \n\\
&+ \left[i\pi\hbar p^0\delta^{(3)}(\mathbf{p}-\mathbf{p}_1)
\left(\langle p_1,p_2|t|p_1,p_2\rangle -\langle p_1,p_2|t^\dagger|p_1,p_2\rangle\right)
+(1\leftrightarrow2) \right] 
\Bigg\}\;, \label{wwww}
\end{align}
where the symbol $(1\leftrightarrow2)$ denotes the exchange of the indices $1$ and $2$,
$dP$  is the abbreviation of $d^4p\, \delta(p^2-m^2)$, and 
\begin{align}
&\langle{p,p^\prime|t|p_1,p_2}\rangle\equiv
\frac{(2\pi\hbar)^{4}}{2\sqrt{\pi}}\,  \outlanglesm p^\prime|:\rho(0):| 
p_1,p_2\inranglesm
 \label{turho}
\end{align}
is the $t$ matrix element  i.e., conventional scattering amplitude derived in \cite{DeGroot:1980dk}  originating from the unspecified interaction $\rho$, which can be 
computed in a  standard way 
via familiar techniques in
quantum field theory.  To obtain a specific form of local collision kernal,
we combine Eq.~\eqref{wwww} with Eq.~\eqref{generalcollisionterm}
and get 
\begin{align}
\label{cloc1}
 C_{22,l}
 = &2\delta(p^2-m^2)\int d^4 p_{1}d^4 p_2
 \Bigg\{ \int dP^\prime 
 \delta(p+p^\prime-p_1-p_2) \langle{p,p^\prime|t|p_1,p_2}\rangle
 \langle{p_1,p_2|t^\dagger|p,p^\prime}\rangle \n\\
 &+ \left[i\pi\hbar p^0\delta^{(3)}(\mathbf{p}-\mathbf{p}_1)
 \left(\langle p_1,p_2|t|p_1,p_2\rangle -\langle p_1,p_2|t^\dagger|p_1,p_2\rangle\right)
 +(1\leftrightarrow2) \right] 
 \Bigg\}
 \prod_{j=1}^2 W(x, p_j)\; ,
 \end{align}
The factor $\delta(p^2-m^2)$ in the front of the r.h.s of $C_{22}$ shows the on-shell property of  the local term as is mensioned several times previously. In deriving the above equation
the formula  $\tilde{\Delta}_R(p)-\tilde{\Delta}_R^\star(p)=2\pi i \hbar^2 \delta (p^2 - m^2)$ is employed with 
\begin{equation}
 \tilde{\Delta}_R(p)=-\frac{\hbar^2}{p^2-m^2+i\epsilon p^0} \;, \label{gapp}
\end{equation}
 appearing in \eq\eqref{offshellscatteringfinal} and taking the form of propogator of Klein-Gordan field.

In order to bridge Wigner function with one-particle distribution function, we  make quasi-particle approximation,
\begin{align}
&W(x,p)= \delta(p^2-M^2)f(x,p)=\delta(p^2-m^2-\hbar\delta m^2)f(x,p)\n\\
&=\delta(p^2-m^2)f(x,p)-\hbar\delta^\prime(p^2-m^2)\delta m^2f(x,p),\label{W0}
\end{align}
where the mass shift $\delta m^2$ is determined from the on-shell condition for $W$ shown by Eq.(\ref{onshellwiggg}) up to order $O(\hbar)$,
\begin{align}
\hbar\delta(p^2-m^2)\delta m^2f(x,p)=\hbar\delta M,
\end{align}
where $x\delta^\prime(x)=-\delta(x)$  is used.

We identify $W^{(0)}$ with the first term of the second line of Eq.(\ref{W0}) and Substitution of $W^{(0)}$ into Eq.(\ref{cloc1}) lead us to
\begin{eqnarray}
 C_{22,l}[f]&=& 
2\int dP_1\, dP_2\, 
 dP^\prime\,   \delta(p^2-m^2)\delta^{(4)}(p+p^\prime-p_1-p_2)\langle{p,p^\prime|t|p_1,p_2}\rangle
  \langle{p,p_2|t^\dagger|p,p^\prime}\rangle
  \prod_{j=1}^2  f(x,p_j)\n\\
  & + &i\, 2\pi\hbar 
  \int  dP_2\, dP_1\,p_0\delta^{(3)}(\mathbf{p}-\mathbf{p}_1)\delta(p^2-m^2)     f(x,p_1)f(x,p_2)
  \langle{p_1,p_2|t-t^\dagger|p_1,p_2}\rangle +(1\leftrightarrow2)  \;.
\end{eqnarray}
Noticing that the term involving 
the amplitude with the operator $t-t^\dagger$ is related to the first term through the 
optical theorem \cite{DeGroot:1980dk}
\begin{equation}
i\pi\hbar\langle p,p_1|t-t^\dagger|p,p_1\rangle 
= - \int dP^\prime dP_1^\prime  \delta^{(4)}(p^\prime+p_1^\prime-p-p_1)
\langle p,p_1|t|p^\prime,p_1^\prime\rangle
\langle p^\prime,p_1^\prime|t^\dagger|p,p_1\rangle \; ,
\label{opttheo}
\end{equation}
which provides the gain and loss part of collision kernel related by detailed equilibrium.

And eventually the local Boltzmann equation is  
\begin{eqnarray}
 \delta(p^2-m^2)p \cdot \partial \, f(x,p)&=&C_{22,l}[f]\;. \label{prop1}
\end{eqnarray}
where the collision term is given by 
\begin{eqnarray}
C_{22,l}[f] &\equiv  &
 \int dP_1\, dP_2\, dP^\prime\,
 \mathcal{W}\big[ f(x,p_1)f(x,p_2) 
 -f(x,p)f(x,p^\prime)\big]\;,
 \label{localcollafter}
\end{eqnarray}
with the collision rate
\begin{eqnarray}
\mathcal{W}&\equiv& 2\delta^{(4)}(p+p^\prime-p_1-p_2) \langle{p,p^\prime|t|p_1,p_2}\rangle
  \langle{p_1,p_2|t^\dagger|p,p^\prime}\rangle \;,
  \label{local_col_GLW_after}
\end{eqnarray}
where the appearance of extra $2$ factor  in transition rate can be absorbed into the definition of $f(x,p)$ noting our definition of $f(x,p)$ is half of the one used in \cite{DeGroot:1980dk}.
This is exactly what we want up to now, namely, the result of the zero order $O(\hbar^0)$ recovers the local Boltzmann equation consistent with our common belief.

\section{Nonlocal collisions for dilute gases}
\label{secnon}

In this section, we are ready to compute the nonlocal collision term, which correspondent to the contribution from the second term in 
the second line of Eq.~\eqref{generalcollisionterm}. Owing to the existence of 
$\partial^\mu_{q_j} \delta^{(4)}(q_j)$ term, the 
momentum $p^\mu$ of the Wigner function may not be on-shell.
To put it clearly, after integration by parts, the nonlocal kernel in 
\eq\eqref{generalcollisionterm} can be cast into  
\begin{align}
 C_{22,nl}^{(1)} =-i\hbar {}&  (2\pi\hbar)^6
\int d^4 p_{1}d^4 p_{2}d^4 q_{1}d^4 q_{2} \, \delta^{(4)}(q_1)\delta^{(4)}(q_2)\n\\
  &\times\left\{  W(x, p_1)
\left( \partial_{\mu}W(x,p_2)\right)
  \partial_{q_2}^\mu+ \left( \partial_{\mu}W(x, p_1) \right)
 W(x,p_2)
\partial_{q_1}^\mu\right\}\n\\
&\times\,   
\inlangle{p_1-\frac{q_1}{2},p_2-\frac{q_2}{2}\Big|\Phi(p)\Big|
p_1+\frac{q_1}{2}, p_2+\frac{q_2}{2}}\inrangle \;,
\label{cnlsecond}
\end{align}
Here we keep our focus on the nonlocal collision term up to the order of $O(\hbar)$ in the scheme of $\hbar$ expansion i.e. gradients expansion. As shown in \eq\eqref{W0}, we need to replace the Wigner function $W$ with the zeroth-order 
distribution function by power counting. Mass shift term $\delta m$ entering the nonlocal 
collision term is at least of second order in gradients. 

With a rather lengthy  calculation and details put in App.~\ref{non_loc_coll_calc}, it is shown  that $C_{nl}^{(1)}$ can 
be divided into an on-shell and an off-shell part,
\begin{align}
C_{22,nl}^{(1)}
={}&C_{\text{22,off-shell}}^{(1)}+ \delta(p^2-m^2)
  C_{22,\os}^{(1)} \;,
\end{align}
and such a division can be also seen in \cite{Weickgenannt:2021cuo}.
Moreover, it is proved that the off-shell contribution $C_{22,\text{off-shell}}^{(1)}$ exactly cancels the 
off-shell part of the left-hand side of the Boltzmann equation \eqref{wigboltz9} 
with $W$ replaced by one particle distribution function  according to \eq\eqref{W0}. 
Last but not the least, the explicit expression for $C_{22,\os}^{(1)}$ contains momentum derivatives of 
matrix elements and is also 
formulated in App.~\ref{non_loc_coll_calc}. 
In \cite{Weickgenannt:2021cuo},
this term is neglected  assuming the 
scattering amplitude to be constant over scales of the order of the interaction range.
Here we keep this term and therefore the Boltzmann equation containing only on-shell contributions 
can be written in terms of the distribution function $f(x,p)$ as
\begin{equation}
 \delta(p^2-m^2)\, p \cdot \partial f(x,p)=\delta(p^2-m^2)\, 
 C_{22,\text{on-shell}}[f]\;, \label{onshellcolll}
\end{equation}
with
\begin{equation}
\label{coll123}
 C_{22,\text{on-shell}}[f]\equiv  C_{22,l}[f]+\hbar\, C_{22,\os}^{(1)}[f] \;,
\end{equation}
where $ C_{22,l}$ is the local term calculated in the previous section.
Compared to the results in \cite{Weickgenannt:2021cuo} after the cancellation of those off-shell parts, we are only left with a term related to momentum derivatives of 
matrix elements for lack of spinor structure. When talking about the collision term of massive spin $1/2$ particles, this extra term plays a big role in constituting an appealing structure of Enskog-type nonlocal shift from collision center. The appearance of this shift results from the interplay of spin  and orbital angular momentum, in other words, spin-orbit coupling. In view of this, the lack of Enskog-type shift reflects the lack of the dynamic effects of spin because we are now only taking spinless particles into account.

\section{dense gases}
\label{dense}
The above derivation for the collision term is confined to two-body scattering, which contribute most to the transport of dilute gases. Although nonlocality is taken into account, it is not enough for constructing a kinetic theory for realistic dense gases. To that end,  the corrections to $W$ must be considered to yield kinetic equations for denser system revealed by the ellipsis in Eq.(\ref{ass1}). 
Thus, for a denser system, we  approximate
\begin{align}
\label{ass3}
W&=W_\text{in}+\frac{1}{2}\int d^4 x^\prime_1 d^4x^\prime_2d^4p^\prime_1d^4p^\prime_2\Psi_2(x^\prime_1,x^\prime_2,p^\prime_1,p^\prime_2|p)W_\text{in}(x+x^\prime_1,p^\prime_1)W_\text{in}(x+x^\prime_2,p^\prime_2),
\end{align} 
and $\Psi$ is defiend as
\begin{align}
\Psi_2(x^\prime_1,x^\prime_2,p^\prime_1,p^\prime_2|p)& \equiv \frac{1}{(2\pi\hbar)^2}
\int d^4 q^\prime_1d^4 q^\prime_2\,e^{iq^\prime_1\cdot x^\prime_1+iq^\prime_2\cdot x^\prime_2}
{\phantom{\Big|}}_\text{in}
\Big\langle{p^\prime_1-\frac{q^\prime_1}{2},p^\prime_2-\frac{q^\prime_2}{2} \Big| \Psi(p) 
	\Big| p^\prime_1+\frac{q^\prime_1}{2}, p^\prime_2+\frac{q^\prime_2}{2}}\Big\rangle_\text{in},
\label{psigen}
\end{align} 
with
\begin{align}
\Psi(p)& \equiv 4\pi\int\frac{d^4y}{(2\pi\hbar)^5} 
e^{-\frac i\hbar p\cdot y} 
\phi^\dagger\left(\frac y2\right)  
\phi\left(-\frac y2\right).
\label{psigen1}
\end{align} 
Inserting the completeness relation for one particle state and  employing Eq.(\ref{evofphi}), we obtain
\begin{align}
\Psi_2(x^\prime_1,x^\prime_2,p^\prime_1,p^\prime_2|p) &=\frac{4\pi}{(2\pi\hbar)^3}
\int\frac{d^3p^\prime}{p^\prime_0}d^4 q^\prime_1d^4 q^\prime_2e^{iq^\prime_1\cdot x^\prime_1+iq^\prime_2\cdot x^\prime_2}\,\delta^{(4)}(p+p^\prime-p^\prime_1-p^\prime_2)\n\\
&\Big(\left[
\delta^{(3)}\left(\mathbf{p^\prime_2}-\frac{\mathbf{q^\prime_2}}{2}-\mathbf{p}^\prime\right)+\delta^{(3)}\left(\mathbf{p^\prime_1}-\frac{\mathbf{q^\prime_1}}{2}-\mathbf{p}^\prime\right)\right]
\frac{p^{\prime0}}{\sqrt{2(2\pi)^3}\hbar}\n\\
&
+  \tilde{\Delta}_R\left(p^\prime-p^\prime_1+\frac{q^\prime_1}{2}-p^\prime_2  +\frac{q^\prime_2}{2}\right)
\inlangle{p^\prime_1-\frac{q^\prime_1}{2},p^\prime_2-\frac{q^\prime_2}{2}}\Big|:\rho^\dagger(0):\Big|
p^\prime\outrangle\Big)\n\\
&\times\Big( \left[
\delta^{(3)}\left(\mathbf{p^\prime_2}+\frac{\mathbf{q^\prime_2}}{2}-\mathbf{p}^\prime\right)+\delta^{(3)}\left(\mathbf{p^\prime_1}+\frac{\mathbf{q^\prime_1}}{2}-\mathbf{p}^\prime\right)\right]
\frac{p^{\prime0}}{\sqrt{2(2\pi)^3}\hbar}\n\\
&+ \tilde{\Delta}^\star_R\left(p^\prime-p^\prime_1-\frac{q^\prime_1}{2}-p^\prime_2  -\frac{q^\prime_2}{2}\right)\outlangle{p^\prime\Big|:\rho(0):\Big|
	p^\prime_1+\frac{q^\prime_1}{2},p^\prime_2+\frac{q^\prime_2}{2}}\inrangle\Big).
\label{psigen24e }
\end{align} 
Iteration of inserting Eq.(\ref{ass3}) into Eq.(\ref{Wcollttt}) leads us to the correction to the collision term $C_{22}$,
\begin{align}
&C_{23}={} \frac{1}{2}
\int d^4 x_{1}d^4 x_2 d^4 x_3 d^4x_4d^4 p_{1}d^4p_2d^4p_3d^4 p_4d^4 q_{1}d^4 q_{2}\n\\
&\quad\,\,\,\,\times\exp\Big(\frac{i}{\hbar}\big(q_1\cdot x_1+\frac{q_2}{2}\cdot (x_2+x_3-x_4)\,\big)\Big)
{\phantom{\Big|}}_\text{in}
\Big\langle{p_1-\frac{q_1}{2},p_4-\frac{q_2}{2} \Big| \Phi(p) 
	\Big| p_1+\frac{q_1}{2}, p_4+\frac{q_2}{2}}\Big\rangle_\text{in}\n\\
&
\quad\,\,\,\,\times\Psi_2\big(\frac{x_2+x_4-x_3}{2},\frac{x_3+x_4-x_2}{2},p_2,p_3|p_4\big)W(x+x_1, p_1)W(x+x_2,p_2)W(x+x_3,p_3)\, ,  \label{Wcollttt1}
\end{align}
where we have neglected the terms of more than three $W$, which belongs to the contribution of  higher order in density.  Because of three $W$ in it, $C_{23}$ can be seen as a part of triple collision term. To seek the complete form of triple collisions, we have to extend  Eq.(\ref{Wcollttt}) to include three-particle states, which is written as
\begin{align}
C_{33}={}& \frac{2}{3}\frac{1}{(2\pi\hbar)^3}
\int d^4 x_{1}d^4 x_{2}d^4 x_{3}d^4 p_{1}d^4 p_{2}d^4 p_{3}d^4 q_{1}d^4 q_{2}d^4 q_{3}\n\\
&
{\phantom{\Big|}}_\text{in}
\Big\langle{p_1-\frac{q_1}{2},p_2-\frac{q_2}{2}, p_3-\frac{q_3}{2}\Big| \Phi(p) 
	\Big| p_1+\frac{q_1}{2}, p_2+\frac{q_2}{2}},p_3+\frac{q_3}{2}\Big\rangle_\text{in}\prod_{j=1}^3 \exp\left(\frac{i}{\hbar}q_j\cdot x_j\right)
W(x+x_j, p_j)\; ,    \label{triple}
\end{align} 
where the substitution of $W_{\text{in}}$ is made with $W$ and the detailed form of in-matrix element for three-particle states is put in App.\ref{collkerapp}. Considering very small phase space for local triple collisions,  we here and now do not seek gradients expansion which effectively separate local and nonlocal parts of collision kernel.

Then the full transport function for dense gases are shown by
\begin{equation}
\delta(p^2-m^2)\, p \cdot \partial f(x,p)=\delta(p^2-m^2)\, 
C_{22,\text{on-shell}}[f]+C_{23}[f]+C_{33}[f]\;, \label{total}
\end{equation}
with Wigner function $W$ within Eqs.(\ref{Wcollttt1}) and (\ref{triple}) replaced by $f$ according to Eq.(\ref{W0}) and $C_{22,\text{on-shell}}[f]$ is on-shell collision term up to $\mathcal{O}(\hbar)$ seeing Eq.(\ref{coll123}). Detailed $\hbar$ power count is not performed here about $C_{23}$ and $C_{33}$. To our knowledge, the triple collision term is first proposed in the thesis of \cite{Choh} by Choh and Uhlenbeck, who develop Bogoliubov's ideas on initial correlation \cite{Bogoliubov}. Such a pioneering work is based on the expansion of particle groups just like Mayer's cluster expansion for the EOS in equilibrium.  In the 1980's, Van Weert and his cooperator develop the related works and obtain the triple collision term appropriate for nonrelativistic quantum statistical in a similar fashion to Degroot's formalism based on quantum field theory.
Here we argue Eq.(\ref{total}) can be seen as the relativistic extension of the results of \cite{Choh, deboerw, deboerv} as these equations share the same structure though with different transition rates. Because the discussion up to now is restricted to spin $0$ particles, the transport equation for particles with other spin remains to be derived, which demands a laborious work considering more complex Lorentz structure for the particle field involved.

There are some comments about this transport equation. Obtained from Wigner formalism and quantum field theory, the on-shell condition for Wigner function or one particle distribution function must be put into consideration simultaneously, which follows the similar footing leading to the transport equation because they share a similar structure. $C_{23}[f]$ denotes the contribution to triple collisions with only  in-matrix element of two-particle states involved, while $C_{33}[f]$ represent also a part of  the triple collision term corresponding to  in-matrix element of three-particle states. Noting that the Wigner function act as the density, such an expansion is called density expansion. That is why we only recover Boltzmann equation in the nondegenerate limit. Keeping the proceture of density expansion to quadruple collisions, the usual Boltzmann equation with Bose enhancement can be recovered shown in \cite{deboer}.  In order not to complicate the discussion about dense gases, we are contend with the triple collision term. For more details and a clear physical picture about triple collisions, we refer the readers to \cite{Dorfman}. On the other hand, the gradients expansion for higher density collision term can be also implemented, which will separate the local and nonlocal contributions explicitly just as mentioned above. Such a calculation and subsequent simplification for the triple collision term will be  our future plan. It is known that the properties of dense gases are well captured by nonlocality, multi collision members, high degenaracy and also off-shell effects. With all those effects or part of those included, the highly involved and complicated collision term shall be attached to great importance when the system is dense enough.



\section{Conclusions} 
\label{secconc}

In this paper, we provide a detailed and systematic derivation of  the relativistic kinetic equation and collision kernel  for dense gases of spin-$0$ particles from quantum field theory based on the Wigner-function formalism. By adopting the scheme of  density expansion, 
the kinetic equation obtained in the lowest order in density recovers the Boltzmann equation with a nonlocal binary collision term.  Keeping density expansion to the next order,  the triple collision term is obtained, which can be seen as a relativistic  extension to the pioneers' works on the  related research about dense gases in nonrelativistic cases. Though derived from the theory of spin-$0$ field, we can also treat the particles in consideration as classical spinless particles when spin induced effects are not so comparably large and thus the kinetic equation works in that case. Without losing generality, we do not specify the form of interaction from beginning to end and transition rates are all expressed in terms of $t$ or $S$ matrix elements that can be evaluated in a systematic way using standard techniques in quantum field theory, however, generally speaking, most of the interactions we met in quantum field theory is transferred by gauge bosons, which means gauge link must be plugged into the definition of Wigner function from the beginning to complete the story. To note, this is also the case in the community of heavy-ion collisions where people show great interests in the phenomena of spin polarization and alignment observed in experiments \cite{Adam:2018ivw, Acharya:2019vpe}. In that case, spin effects come back into play demanding a more complicated computation. We do not intend to fix the issue of non-gauge  in this paper and leave it to a future work. Noting there are many phenomena tightly related to dense transport in collider physics, astrophysics, and cosmoslogy, the equation obtained here can be put into a pratical use.
On the other hand, with this equation at hand,  the corrections to transport coefficients can be sought taking the form of density expansion like collision kernel, which of the form can be found in \cite{Dorfman, Choh, Eu}.  Since there are many efforts following Uhlenbeck-Choh theory listed above, their practices can be taken as the  reference for the extension to relativistic cases, which is also in our future plan.

\section*{Acknowledgments}
J.H. is grateful to Jiaxing Zhao for reading the script.  This work was supported by the NSFC Grant No.11890710 and No.11890712.

\begin{appendix}

\section{Calculation of the expectation value of $\Phi$} \label{collkerapp}

In the main text, we have met several in-matrix elements which needs to be handled with caution, and we will  explicitly compute these matrix elements in this appendix.

First, we look at Eq.~\eqref{generalcollisionterm},
\begin{equation}
\inlangle{p_1-\frac{q_1}{2},p_2-\frac{q_2}{2}\Big|\Phi(p)\Big|
p_1+\frac{q_1}{2}, p_2+\frac{q_2}{2}}\inrangle \;, 
\end{equation}
with the operator $\Phi(p)$  given by \eq\eqref{phigen}.
A completeness basis of free out-states is inserted into the above equation and  we obtain following similar steps  in \cite{DeGroot:1980dk},
\begin{eqnarray}
&&\inlangle{p_1-\frac{q_1}{2},p_2-\frac{q_2}{2}\Big|\Phi(p)\Big|
p_1+\frac{q_1}{2}, p_2+\frac{q_2}{2}}\inrangle\n\\
&=& i \int dP^\prime \delta^{(4)}(p+p^\prime-p_1-p_2)\label{phi1111} 
\left\{\outlangle{p^\prime\Big|:\rho(0):\Big|
	p_1+\frac{q_1}{2}, p_2+\frac{q_2}{2}}\inrangle 
\inlangle{p_1-\frac{q_1}{2},p_2-\frac{q_2}{2}\Big|\phi^\dagger(0)\Big|
	p^\prime}\outrangle\right.\n\\
& &\,\left.-\outlangle{p^\prime\Big|\phi(0)\Big|
	p_1+\frac{q_1}{2}, p_2+\frac{q_2}{2}}\inrangle 
\inlangle{p_1-\frac{q_1}{2},p_2-\frac{q_2}{2}\Big|:\rho^\dagger(0):\Big|
	p^\prime}\outrangle
 \right\} \;,
\n 
\end{eqnarray}
where  the fact that one- and two-particle 
states are all eigenstates of the total momentum operator is used,  meanwhile, the expectation value of the operator
 $\phi(-y/2)$ in Heisenberg picture is given by 
\begin{eqnarray}
\label{evofphi}
 \outlangle{p^\prime\Big|\phi\left(-\frac y2\right)\Big|
 p_1+\frac{q_1}{2}, p_2+\frac{q_2}{2}}\inrangle 
 &=& e^{-\frac{i}{2\hbar}(p^\prime-p_1-q_1/2 -p_2-q_2/2)\cdot y}
 \outlangle{p^\prime\Big|\phi(0)\Big|
 p_1+\frac{q_1}{2}, p_2+\frac{q_2}{2}}\inrangle  \;. \n\\
\end{eqnarray}
Invoking Yang-Feldman equation
\begin{equation}
\phi^\dagger (0) = \phi^\dagger_\text{in}(0) + \int d^4 x \, \Delta_R (-x) \rho(x) \;,
\end{equation}
where $\phi_\text{in}$ is defined by
\begin{align}
\phi^\dagger_\text{in}(0)=\frac{1}{\sqrt{2(2\pi)^3}\hbar}\int\frac{d^3p}{p^0}a^\dagger_{\text{in}}(p)
\end{align}
 and $\Delta_R(x)$ is the retarded 
Green's function, 
\begin{equation}
\Delta_R (x) = \frac{1}{(2\pi\hbar)^4} \int d^4p \, \tilde{\Delta}_R(p) e^{- \frac{i}{\hbar} p \cdot x}\;,
\end{equation}
with its Fourier transform
\begin{equation}
 \tilde{\Delta}_R(p)\equiv-\frac{\hbar^2}{p^2-m^2+i\epsilon p^0} \;, \label{stildeapp}
\end{equation}
Then the  matrix element of $\phi$  and $\phi^\dagger$ are  shown by
\begin{eqnarray}
 \inlangle{p_1-\frac{q_1}{2}, p_2-\frac{q_2}{2}}\Big|\phi^\dagger(0)\Big|p^\prime\outrangle
&= & \left[
\delta^{(3)}\left(\mathbf{p_2}-\frac{\mathbf{q_2}}{2}-\mathbf{p}^\prime\right)+\delta^{(3)}\left(\mathbf{p_1}-\frac{\mathbf{q_1}}{2}-\mathbf{p}^\prime\right)\right]
 \frac{p^{\prime0}}{\sqrt{2(2\pi)^3}\hbar}\label{phi222} \\
 & + & \tilde{\Delta}_R\left(p^\prime-p_1+\frac{q_1}{2}-p_2  +\frac{q_2}{2}\right)
 \inlangle{p_1-\frac{q_1}{2},p_2-\frac{q_2}{2}}\Big|:\rho^\dagger(0):\Big|
 p^\prime\outrangle \;,
\n\\
\outlangle{p^\prime\Big|\phi(0)\Big|
	p_1+\frac{q_1}{2}, p_2+\frac{q_2}{2}}\inrangle
&= & \left[
\delta^{(3)}\left(\mathbf{p_2}+\frac{\mathbf{q_2}}{2}-\mathbf{p}^\prime\right)+\delta^{(3)}\left(\mathbf{p_1}+\frac{\mathbf{q_1}}{2}-\mathbf{p}^\prime\right)\right]
\frac{p^{\prime0}}{\sqrt{2(2\pi)^3}\hbar}\label{phi223} \\
& + & \tilde{\Delta}^\star_R\left(p^\prime-p_1-\frac{q_1}{2}-p_2  -\frac{q_2}{2}\right)
\outlangle{p^\prime\Big|:\rho(0):\Big|
p_1+\frac{q_1}{2},p_2+\frac{q_2}{2}}\inrangle \;,
\n
\end{eqnarray}
with the conventional orthogonality condition
$\langle p | p^\prime \rangle = p^0 \delta^{(3)}({\bf p}-{\bf p^\prime})
$ used. 
Plutting \eq\eqref{phi222}  , \eqref{phi223},  \eqref{phi1111} , \eqref{stildeapp} and 
\eqref{gapp} altogether,
we obtain
\begin{eqnarray} \label{offshellscattering}
&&\inlangle{p_1-\frac{q_1}{2},p_2-\frac{q_2}{2}\Big|\Phi(p)\Big|
p_1+\frac{q_1}{2}, p_2 +\frac{q_2}{2}}\inrangle\n\\
&=&\frac i2\bigg\{\frac{1}{\sqrt{2(2\pi)^3}\hbar}
\left[\delta^{(3)}\left(\mathbf{p}-\mathbf{p_1}
-\frac{\mathbf{q_2}}{2}\right)\delta\left(p^0+\sqrt{\left(\mathbf{p_2}
-\frac{\mathbf{q_2}}{2}\right)^2+m^2}-E_{p_1}-E_{p_2}\right)\right.\bigg.\n\\
&&\times\left.\bigg.\outlangle{p_2-\frac{q_2}{2}\Big|:\rho(0):\Big|
	p_1+\frac{q_1}{2}, p_2+\frac{q_2}{2}}\inrangle+(1\leftrightarrow2)\bigg. \right]\n\\
&&- \bigg.\frac{1}{\sqrt{2(2\pi)^3\hbar}}\left[\delta^{(3)}\left(\mathbf{p}-\mathbf{p_1}
+\frac{\mathbf{q_2}}{2}\right)
\delta\left(p^0+\sqrt{\left(\mathbf{p_2}+\frac{\mathbf{q_2}}{2}\right)^2+m^2}
-E_{p_1}-E_{p_2}\right)\right.\bigg.\n\\
&&\times\left.
\inlangle{p_1-\frac{q_1}{2},p_2-\frac{q_2}{2}\Big|:\rho^\dagger(0):\Big|p_2+\frac{q_2}{2}}\outrangle 
+(1\leftrightarrow2)\right]\bigg.\n\\
&&\bigg.+2\int dP^\prime\delta^{(4)}(p+p^\prime-p_1-p_2)
\left[\tilde{\Delta}_R\left(-p+\frac{q_1+q_2}{2}\right)-\tilde{\Delta}^\star_R\left(-p-\frac{q_1+q_2}{2}\right)\right]
\bigg.\n\\
&&\times \,
\outlangle{p^\prime\Big|:\rho(0):\Big|p_1
+\frac{q_1}{2}, p_2+\frac{q_2}{2}}\inrangle\bigg.\bigg.\inlangle{p_1-\frac{q_1}{2},p_2-\frac{q_2}{2}\Big|:\rho^\dagger(0):\Big|
p^\prime}
\outrangle \bigg\} \;,
\end{eqnarray}
with the notation $E_p\equiv \sqrt{\mathbf{p}^2+m^2}$. 
Eventually, we make use of  Eq.~\eqref{turho} to write
\begin{align} \label{offshellscatteringfinal}
&w(p_1,q_1,p_2,q_2,p)= \inlangle{p_1-\frac{q_1}{2},p_2-\frac{q_2}{2}\Big|\Phi(p)\Big|p_1+\frac{q_1}{2}, p_2
 +\frac{q_2}{2}}\inrangle\n\\
 &= \frac{1}{2(2\pi\hbar)^6} \Big(2 \int dP^\prime\, 
 \frac{i}{\pi\hbar^2}\left[ \tilde{\Delta}_R\left(-p+\frac{q_1+q_2}{2}\right)-\tilde{\Delta}^\star_R\left(-p-\frac{q_1+q_2}{2}\right)\right]\n\\
 &\times\delta^{(4)}(p+p^\prime-p_1-p_1)\Big\langle{p+\frac{q_1+q_2}{2}, 
 p^\prime\Big|t\Big|p_1+\frac{q_1}{2}, p_2+\frac{q_2}{2}}\Big\rangle\n\\
 &\times\Big\langle{p_1-\frac{q_1}{2},p_2-\frac{q_2}{2}\Big|t^\dagger\Big|
 p-\frac{q_1+q_2}{2},p^\prime}\Big\rangle \n\\
  &+i2\pi\hbar\delta^{(3)}\left(\mathbf{p}-\mathbf{p_1}-\frac{\mathbf{q_2}}{2}\right)
  \delta\left(p^0+p_2^0-\frac{q_2^0}{2}-E_{p_1}-E_{p_2}\right)\n\\
 &\times\Big\langle{p_1+q_2+\frac{q_1}{2},p_2-\frac{q_2}{2}\Big|t\Big|
 p_1+\frac{q_1}{2},p_2+\frac{q_2}{2}}\Big\rangle+(1\leftrightarrow2)\n\\
 &-i2\pi\hbar \delta^{(3)}\left(\mathbf{p}-\mathbf{p_1}+\frac{\mathbf{q_2}}{2}\right)
 \delta\left(p^0+p_2^0+\frac{q_2^0}{2}-E_{p_1}-E_{p_2}\right)\n\\
 &\times\Big\langle{p_1-\frac{q_1}{2},p_2-\frac{q_2}{2}\Big|t^\dagger\Big|
 p_1-q_2-\frac{q_1}{2},p_2+\frac{q_2}{2}}\Big\rangle+(1\leftrightarrow2)\,\Big)\; ,
\end{align}
where  to linear order in $\mathbf{q}_2$ we may replace 
$\sqrt{(\mathbf{p_2}\pm \mathbf{q_2}/2)^2+m^2}= p_2^0\pm q_2^0/2$.  Because we only consider zeroth and first order terms of a Taylor expansion in $q_i$ in 
Eq.\ (\ref{generalcollisionterm}) such an approximation is
sufficient.

At the meantime the  in-matrix element for three-particle states is also needed,
\begin{eqnarray}
&&{\phantom{\Big|}}_\text{in}
\Big\langle{p_1-\frac{q_1}{2},p_2-\frac{q_2}{2}, p_3-\frac{q_3}{2}\Big| \Phi(p) 
	\Big| p_1+\frac{q_1}{2}, p_2+\frac{q_2}{2}},p_3+\frac{q_3}{2}\Big\rangle_\text{in}\n\\
&&=2 i \int dP_1^\prime dP_2^\prime \delta^{(4)}(p+p_1^\prime+p_2^\prime-p_1-p_2-p_3)\n\\
&&\quad
\times\left\{\outlangle{p_1^\prime,p_2^\prime\Big|:\rho(0):\Big|
	p_1+\frac{q_1}{2}, p_2+\frac{q_2}{2},p_3+\frac{q_3}{2}}\inrangle 
\inlangle{p_1-\frac{q_1}{2},p_2-\frac{q_2}{2},p_3-\frac{q_3}{2}\Big|\phi^\dagger(0)\Big|
	p_1^\prime,p_2^\prime}\outrangle\right.\n\\
& &\quad\,\,\,\left.-\outlangle{p_1^\prime,p_2^\prime\Big|\phi(0)\Big|
	p_1+\frac{q_1}{2}, p_2+\frac{q_2}{2},p_3+\frac{q_3}{2}}\inrangle 
\inlangle{p_1-\frac{q_1}{2},p_2-\frac{q_2}{2},p_3-\frac{q_3}{2}\Big|:\rho^\dagger(0):\Big|
	p_1^\prime,p_2^\prime}\outrangle\label{phi3} 
\right\} \;.
\end{eqnarray}
The matrix element of $\phi^\dagger$ is thus given by
\begin{eqnarray}
\inlangle{p_1-\frac{q_1}{2}, p_2-\frac{q_2}{2},p_3-\frac{q_3}{2}}\Big|\phi^\dagger(0)\Big|p_1^\prime,p_2^\prime\outrangle
&=& \frac{1}{\sqrt{2(2\pi)^3}\hbar} \left[
\inlangle{ p_2-\frac{q_2}{2},p_3-\frac{q_3}{2}}\Big|p_1^\prime,p_2^\prime\outrangle\right.\n\\
&&\left.+\inlangle{p_1-\frac{q_1}{2}, p_3-\frac{q_3}{2}}\Big|p_1^\prime,p_2^\prime\outrangle+\inlangle{p_1-\frac{q_1}{2}, p_2-\frac{q_2}{2}}\Big|p_1^\prime,p_2^\prime\outrangle\right]\n \\
&& +  \tilde{\Delta}_R\left(p_1^\prime+p_2^\prime-p_1+\frac{q_1}{2}-p_2  +\frac{q_2}{2}-p_3+\frac{q_3}{2}\right)\n\\
&&\times\inlangle{p_1-\frac{q_1}{2},p_2-\frac{q_2}{2},p_3-\frac{q_3}{2}}\Big|:\rho^\dagger(0):\Big|
p_1^\prime,p_2^\prime\outrangle \;
\end{eqnarray}
and the following reduction formula is helpful 
\begin{align}
& \outlangle p_1^\prime,p_2^\prime,p_3^\prime\big|
p_1,p_2,p_3\inrangle=\outlangle p_1^\prime,p_2^\prime\big|a_{\text{in}}(p_3^\prime)\big|
p_1,p_2,p_3\inrangle+i\int d^4xf^\star_{p_3^\prime}(x)\outlangle p_1^\prime,p_2^\prime\big|:\rho(x):\big|
p_1,p_2,p_3\inrangle\, \n\\
&\quad\quad\quad\quad=\outlangle p_1^\prime,p_2^\prime\big|a_{\text{in}}(p_3^\prime)\big|
p_1,p_2,p_3\inrangle+i\frac{(2\pi\hbar)^3}{2\sqrt{\pi}}\delta^{(4)}(p_1^\prime+p_2^\prime+p_3^\prime-p_1-p_2-p_3)\outlangle p_1^\prime,p_2^\prime\big|:\rho(0):\big|
p_1,p_2,p_3\inrangle\, .\label{reduct}
\end{align}
This reduction formula Eq.(\ref{reduct}) can be further simplified by using
\begin{align}
\label{3}
& \outlangle p_1^\prime,p_2^\prime,p_3^\prime\big|
p_1,p_2,p_3\inrangle=\inlangle p_1^\prime,p_2^\prime,p_3^\prime\big|
p_1,p_2,p_3\inrangle+\frac{i}{2\pi\hbar}\delta^{(4)}(p_1^\prime+p_2^\prime+p_3^\prime-p_1-p_2-p_3)\langle p_1^\prime,p_2^\prime,p_3^\prime\big|t\big|
p_1,p_2,p_3\rangle\,\n\\
\end{align}
and
\begin{align}
\label{2}
&\outlangle p_1^\prime,p_2^\prime\big|a_{\text{in}}(p_3^\prime)\big|
p_1,p_2,p_3\inrangle=p^0_1\delta(\mathbf{p^\prime_3}-\mathbf{p_1})\outlangle p_1^\prime,p_2^\prime\big|
p_2,p_3\inrangle+p^0_3\delta(\mathbf{p^\prime_3}-\mathbf{p_3})\outlangle p_1^\prime,p_2^\prime\big|
p_1,p_2\inrangle\n\\
&\quad\quad\quad\quad\quad\quad\quad\quad\quad\quad\quad\quad\quad\,\quad+p^0_2\delta(\mathbf{p^\prime_3}-\mathbf{p_2})\outlangle p_1^\prime,p_2^\prime\big|
p_1,p_3\inrangle\n\\
&\quad\quad\quad\quad\quad\quad\quad\quad\quad\quad\quad\quad\quad\,=p^0_1\delta(\mathbf{p^\prime_3}-\mathbf{p_1})\inlangle p_1^\prime,p_2^\prime\big|
p_2,p_3\inrangle+p^0_3\delta(\mathbf{p^\prime_3}-\mathbf{p_3})\inlangle p_1^\prime,p_2^\prime\big|
p_1,p_2\inrangle\n\\
&\quad\quad\quad\quad\quad\quad\quad\quad\quad\quad\quad\quad\quad\,\quad+p^0_2\delta(\mathbf{p^\prime_3}-\mathbf{p_2})\inlangle p_1^\prime,p_2^\prime\big|
p_1,p_3\inrangle\n\\
&\quad\quad\quad\quad\quad\quad\quad\quad\quad\quad\quad\quad\quad\,\quad+\frac{i}{2\pi\hbar}p^0_1\delta(\mathbf{p^\prime_3}-\mathbf{p_1})\delta^{(4)}(p_1^\prime+p_2^\prime-p_2-p_3)\langle p_1^\prime,p_2^\prime\big|t\big|
p_2,p_3\rangle\n\\
&\quad\quad\quad\quad\quad\quad\quad\quad\quad\quad\quad\quad\quad\,\quad+\frac{i}{2\pi\hbar}p^0_2\delta(\mathbf{p^\prime_3}-\mathbf{p_2})\delta^{(4)}(p_1^\prime+p_2^\prime-p_1-p_3)\langle p_1^\prime,p_2^\prime\big|t\big|
p_1,p_3\rangle\n\\
&\quad\quad\quad\quad\quad\quad\quad\quad\quad\quad\quad\quad\quad\,\quad+ \frac{i}{2\pi\hbar}p^0_3\delta(\mathbf{p^\prime_3}-\mathbf{p_3})\delta^{(4)}(p_1^\prime+p_2^\prime-p_1-p_2)\langle p_1^\prime,p_2^\prime\big|t\big|
p_1,p_2\rangle .
\end{align}
With the detailed expression given in \cite{DeGroot:1980dk}, we can verify that the first term in Eq.(\ref{3}) cancels the sum of the first three term of behind the last equal sign of Eq.(\ref{2}). Thus attention is only paid to other terms within these two equations.
Furthermore, we argue that $S$ matrix elements should have a factor for $4$ momentum conservation as it should be \cite{weinberg}. Because  this factor is already in front of $t$ matrix elements, thus the identity matrix elements should also include $\delta^{(4)}(p_i-p_f)$, and we replace $p^{0}_i\delta^{(3)}(p_i-p_f)\delta(p_i^2-m^2)$ with $\frac{1}{2}\delta^{(4)}(p_i-p_f)$, where the on-shell condition for the asymptotic particle states is explicitly added,
thus
\begin{align}
& \frac{(2\pi\hbar)^3}{2\sqrt{\pi}}\outlangle p_1^\prime,p_2^\prime\big|:\rho(0):\big|
p_1,p_2,p_3\inrangle=\frac{1}{2\pi\hbar}\Big(\langle p_1^\prime,p_2^\prime,p_3^\prime\big|t\big|
p_1,p_2,p_3\rangle-\delta^{(4)}(p_1^\prime+p_2^\prime-p^\prime_3+p_1-p_2-p_3)\langle p_1^\prime,p_2^\prime\big|t\big|
p_2,p_3\rangle\n\\
&\quad\quad\quad\quad-\delta^{(4)}(p_1^\prime+p_2^\prime-p^\prime_3+p_2-p_1-p_3)\langle p_1^\prime,p_2^\prime\big|t\big|
p_1,p_3\rangle- \delta^{(4)}(p_1^\prime+p_2^\prime-p^\prime_3+p_3-p_1-p_2)\langle p_1^\prime,p_2^\prime\big|t\big|
p_1,p_2\rangle\, \Big),\n\\\label{red}
\end{align}
where $\delta(a)\delta(b)=2\delta(a+b)\delta(a-b)$ is used.

Substitution of Eq.(\ref{red}) and its Hermite congugation into Eq.(\ref{phi3}) gives
\begin{eqnarray} \label{offshellscattering1}
&&\Big\langle{p_1-\frac{q_1}{2},p_2-\frac{q_2}{2}, p_3-\frac{q_3}{2}\Big| \Phi(p) 
	\Big| p_1+\frac{q_1}{2}, p_2+\frac{q_2}{2}},p_3+\frac{q_3}{2}\Big\rangle_\text{in}\n\\
&=& 2i\bigg\{\frac{1}{\sqrt{2(2\pi)^3}\hbar}\int dP_1^\prime dP_2^\prime \delta^{(4)}(p+p_1^\prime+p_2^\prime-p_1-p_2-p_3)\n\\
&&
\times\left[\Big(\,\inlangle{ p_2-\frac{q_2}{2},p_3-\frac{q_3}{2}}\Big|p_1^\prime,p_2^\prime\outrangle+\inlangle{p_1-\frac{q_1}{2}, p_3-\frac{q_3}{2}}\Big|p_1^\prime,p_2^\prime\outrangle\right.\bigg.\n\\
&&+\inlangle{p_1-\frac{q_1}{2}, p_2-\frac{q_2}{2}}\Big|p_1^\prime,p_2^\prime\outrangle\Big)\left.\bigg.\outlangle{p_1^\prime,p_2^\prime\Big|:\rho(0):\Big|
	p_1+\frac{q_1}{2}, p_2+\frac{q_2}{2},p_3+\frac{q_3}{2}}\inrangle \bigg. \right]\n\\
&&- \bigg.\frac{1}{\sqrt{2(2\pi)^3\hbar}}\int dP_1^\prime dP_2^\prime \delta^{(4)}(p+p_1^\prime+p_2^\prime-p_1-p_2-p_3)\n\\
&&
\times\left[\Big(\,\outlangle{p_1^\prime,p_2^\prime \Big|p_2+\frac{q_2}{2},p_3+\frac{q_3}{2}}\inrangle+\outlangle{p_1^\prime,p_2^\prime \Big|p_1+\frac{q_1}{2},p_3+\frac{q_3}{2}}\inrangle\right.\bigg.\n\\
&&+\outlangle{p_1^\prime,p_2^\prime \Big|p_1+\frac{q_1}{2},p_2+\frac{q_2}{2}}\inrangle\Big)\left.
\inlangle{p_1-\frac{q_1}{2},p_2-\frac{q_2}{2},p_3-\frac{q_3}{2}\Big|:\rho^\dagger(0):\Big|p^\prime_1,p^\prime_2}\outrangle \right]\bigg.\n\\
&&\bigg.+\int dP_1^\prime dP_2^\prime \delta^{(4)}(p+p_1^\prime+p_2^\prime-p_1-p_2-p_3)\n\\
&&
\times\left[ \tilde{\Delta}_R\left(p_1^\prime+p_2^\prime-p_1+\frac{q_1}{2}-p_2  +\frac{q_2}{2}-p_3+\frac{q_3}{2}\right)-\tilde{\Delta}^\star_R\left(p_1^\prime+p_2^\prime-p_1-\frac{q_1}{2}-p_2  -\frac{q_2}{2}-p_3-\frac{q_3}{2}\right)\right]
\bigg.\n\\
&&\times \,
\outlangle{p_1^\prime,p_2^\prime\Big|:\rho(0):\Big|
	p_1+\frac{q_1}{2}, p_2+\frac{q_2}{2},p_3+\frac{q_3}{2}}\inrangle\bigg.\bigg.\inlangle{p_1-\frac{q_1}{2},p_2-\frac{q_2}{2},p_3-\frac{q_3}{2}\Big|:\rho^\dagger(0):\Big|
	p_1^\prime,p_2^\prime}\outrangle \bigg\} \;,\label{long}
\end{eqnarray}
with
\begin{align}
& \outlangle p_1^\prime,p_2^\prime\big|:\rho(0):\big|
p_1+\frac{q_1}{2}, p_2+\frac{q_2}{2},p_3+\frac{q_3}{2}\inrangle\equiv\frac{2\sqrt{\pi}}{(2\pi\hbar)^4}\Big(\,\langle p_1^\prime,p_2^\prime,p+\frac{q_1}{2}+\frac{q_2}{2}+\frac{q_3}{2}\big|t\big|
p_1+\frac{q_1}{2}, p_2+\frac{q_2}{2},p_3+\frac{q_3}{2}\rangle\n\\
&\quad\quad\quad\quad\quad\quad\quad\quad\quad\quad\quad\quad\quad\quad\quad\quad-\delta^{(4)}(p_1^\prime+p_2^\prime-p-\frac{q_1}{2}-\frac{q_2}{2}-\frac{q_3}{2}+p_1-p_2-p_3)\langle p_1^\prime,p_2^\prime\big|t\big|
p_2+\frac{q_2}{2},p_3+\frac{q_3}{2}\rangle\n\\
&\quad\quad\quad\quad\quad\quad\quad\quad\quad\quad\quad\quad\quad\quad\quad\quad-\delta^{(4)}(p_1^\prime+p_2^\prime-p-\frac{q_1}{2}-\frac{q_2}{2}-\frac{q_3}{2}+p_2-p_1-p_3)\langle p_1^\prime,p_2^\prime\big|t\big|
p_1+\frac{q_1}{2},p_3+\frac{q_3}{2}\rangle\n\\
&\quad\quad\quad\quad\quad\quad\quad\quad\quad\quad\quad\quad\quad\quad\quad\quad- \delta^{(4)}(p_1^\prime+p_2^\prime-p-\frac{q_1}{2}-\frac{q_2}{2}-\frac{q_3}{2}+p_3-p_1-p_2)\langle p_1^\prime,p_2^\prime\big|t\big|
p_1+\frac{q_1}{2},p_2+\frac{q_2}{2}\rangle\,\Big)\, .\n\\\label{n1}
\end{align}
To avoid lengthy representation, we do not further insert Eqs.(\ref{n1})  into Eq.(\ref{long}).


\section{Calculation of nonlocal collision term for dilute gases} \label{non_loc_coll_calc}

Following \cite{Weickgenannt:2021cuo}, this section serves as a derivation of  the nonlocal term for dilute gases. Without causing confusions, we keep the notations used in \cite{Weickgenannt:2021cuo}.  The nonlocal collision term appearing in \eq\eqref{cnlsecond} is given by 
\begin{align} \label{secondcontributiontofirstordercollision}
 C_{22,nl}^{(1)} =-i\hbar {}&  (2\pi\hbar)^6
 \int d^4 p_{1}d^4 p_{2}d^4 q_{1}d^4 q_{2} \, \delta^{(4)}(q_1)\delta^{(4)}(q_2)\n\\
 &\times\left\{  W(x, p_1)
 \left( \partial_{\mu}W(x,p_2)\right)
 \partial_{q_2}^\mu+ \left( \partial_{\mu}W(x, p_1) \right)
 W(x,p_2)
 \partial_{q_1}^\mu\right\}\n\\
 &\times\,   
 \inlangle{p_1-\frac{q_1}{2},p_2-\frac{q_2}{2}\Big|\Phi(p)\Big|
 	p_1+\frac{q_1}{2}, p_2+\frac{q_2}{2}}\inrangle\n\\
 ={}
  &-i\hbar (2\pi\hbar)^6\int d^4p_1 d^4p_2\left\{ \left[\partial_{q_1}^\mu 
  w(p_1,q_1,p_2,q_2,p)  \right]_{q_1=q_2=0}
  W(x,p_2)\partial_{\mu}W(x,p_1)\right. \n\\
  &+\left. \left[\partial_{q_2}^\mu 
  w(p_1,q_1,p_2,q_2,p)\right]_{q_1=q_2=0}
 W(x,p_1)
  \partial_{\mu}W(x,p_2)\right\}\;,
\end{align}

We split $w(p_1,q_1,p_2,q_2,p)$, 
cf.\ \eq\eqref{offshellscatteringfinal}, into a gain term,
\begin{eqnarray}
\lefteqn{\hspace*{-1.0cm} w_{\text{gain}}(p_1,q_1,p_2,q_2,p)=
\frac{i}{\pi\hbar^2(2\pi\hbar)^6}\int dP^\prime\, \left[ \tilde{\Delta}_R\left(-p+\frac{q_1+q_2}{2}\right)-\tilde{\Delta}^\star_R\left(-p-\frac{q_1+q_2}{2}\right)\right]}\n\\
 &\times&\delta^{(4)}(p+p^\prime-p_1-p_1)
 \Big\langle{p+\frac{q_1+q_2}{2},p^\prime\Big|t\Big|
 p_1+\frac{q_1}{2}, p_2+\frac{q_2}{2}}\Big\rangle
 \Big\langle{p_1-\frac{q_1}{2},p_2-\frac{q_2}{2}\Big|t^\dagger\Big|
 p-\frac{q_1+q_2}{2},p^\prime}\Big\rangle \;, \label{gain}
\end{eqnarray}
and a loss term
\begin{eqnarray}
w_{\text{loss}}(p_1,q_1,p_2,q_2,p)&=\frac{1}{2(2\pi\hbar)^6} \Big(i2\pi\hbar\delta^{(3)}\left(\mathbf{p}-\mathbf{p_1}-\frac{\mathbf{q_2}}{2}\right)
\delta\left(p^0+p_2^0-\frac{q_2^0}{2}-E_{p_1}-E_{p_2}\right)\n\\
&\times\Big\langle{p_1+q_2+\frac{q_1}{2},p_2-\frac{q_2}{2}\Big|t\Big|
	p_1+\frac{q_1}{2},p_2+\frac{q_2}{2}}\Big\rangle+(1\leftrightarrow2)\n\\
&-i2\pi\hbar \delta^{(3)}\left(\mathbf{p}-\mathbf{p_1}+\frac{\mathbf{q_2}}{2}\right)
\delta\left(p^0+p_2^0+\frac{q_2^0}{2}-E_{p_1}-E_{p_2}\right)\n\\
&\times\Big\langle{p_1-\frac{q_1}{2},p_2-\frac{q_2}{2}\Big|t^\dagger\Big|
	p_1-q_2-\frac{q_1}{2},p_2+\frac{q_2}{2}}\Big\rangle+(1\leftrightarrow2)\Big)\;. \label{loss}
\end{eqnarray}
Since we are now only interested in the contribution of order $\mathcal{O}(\hbar)$, the
Wigner functions in Eq.\ (\ref{secondcontributiontofirstordercollision}) 
can  be replaced by their
zeroth-order expression, thus the terms $\sim W \partial_\mu W$ 
is also replaced by
$\sim f \partial_\mu f$ view Eq.(\ref{W0}). 
Perform the $q_j$-derivative on the gain part, Eq.\ (\ref{gain}), and the respective terms
in Eq.\  (\ref{secondcontributiontofirstordercollision}) lead to contributions of the form with extra $\delta(p_i^2-m^2)$ integrated into $dP_i$
\begin{align}
 &\left[f(x,p_2)\partial^\mu f(x,p_1)\partial_{q_1\mu}
 +f(x,p_1)\partial^\mu f(x,p_2)\partial_{q_2\mu}\right] 
  \times \left[ \tilde{\Delta}_R\left(-p+\frac{q_1+q_2}{2}\right)
-\tilde{\Delta}_R^\star\left(-p-\frac{q_1+q_2}{2}\right)\right]_{q_1=q_2=0}\n\\
 &= -\frac12\hbar^2\partial^\mu\big( f(x,p_1)f(x,p_2) \big)
 \partial_{q\mu}\left[\frac{1}{(-p+q)^2-m^2+i\epsilon(-p^0+q^0)}
 -\frac{1}{(p+q)^2-m^2+i\epsilon(p^0+q^0)}\right]_{q=0}\n\\
&=\partial^\mu \big(f(x,p_1)f(x,p_2)\big)\frac{p_\mu}{p^2-m^2}
\big(\,\tilde{\Delta}_R(p) + \tilde{\Delta}_R^\star(p)\,\big)
 \;. \label{firstoffshellcontributionfrominteractions}
\end{align}
This is indeed an off-shell contribution 
to the Boltzmann equation for the appearance of the factor $p^2-m^2$ in the denominator.
The  off-shell contributions can also be seen in the $q_j$-derivatives act on the 
loss term, Eq.\ (\ref{loss}), i.e., we write this  in explicit form
\begin{eqnarray}
 &&-\frac{i\hbar}{2}\int dP_1 dP_2\, 
 \left[f(x,p_2)\partial_{\nu}f(x,p_1)\partial_{q_1}^\nu
 +f(x,p_1)\partial_{\nu}f(x,p_2)\partial_{q_2}^\nu\right]\n\\
 &&\times 
 \Big[i2\pi\hbar\delta^{(3)}\left(\mathbf{p}-\mathbf{p_1}-\frac{\mathbf{q_2}}{2}\right)
 \delta\left(p^0+p_2^0-\frac{q_2^0}{2}-E_{p_1}-E_{p_2}\right)\n\\
 &&\times\Big\langle{p_1+q_2+\frac{q_1}{2},p_2-\frac{q_2}{2}\Big|t\Big|
 	p_1+\frac{q_1}{2},p_2+\frac{q_2}{2}}\Big\rangle+(1\leftrightarrow2)\n\\
 &&-i2\pi\hbar \delta^{(3)}\left(\mathbf{p}-\mathbf{p_1}+\frac{\mathbf{q_2}}{2}\right)
 \delta\left(p^0+p_2^0+\frac{q_2^0}{2}-E_{p_1}-E_{p_2}\right)\n\\
 &&\times\Big\langle{p_1-\frac{q_1}{2},p_2-\frac{q_2}{2}\Big|t^\dagger\Big|
 	p_1-q_2-\frac{q_1}{2},p_2+\frac{q_2}{2}}\Big\rangle+(1\leftrightarrow2)\Big]_{q_1=q_2=0}\n\\
 &=& -\frac{i\hbar}{2}\int dP_{2} 
 {\partial_{\nu} \left[f(x,p_2)f(x,p_1)\right]}
 \left(\partial_{q_1}^\nu+\partial_{q_2}^\nu\right)\n\\
 &&\times \Big[- {\frac{i\pi\hbar}{E_{p+\frac{q_2}{2}}}} \delta\left(p^0+\frac{q_2^0}{2}
 -E_{p+\frac{q_2}{2}}\right)
 \Big\langle{p-\frac{q_1-q_2}{2},p_2-\frac{q_2}{2}\Big|t^\dagger\Big|
 p-\frac{q_1+q_2}{2},p_2+\frac{q_2}{2}}\Big\rangle \n\\
 &&+{\frac{i\pi\hbar}{E_{p-\frac{q_2}{2}}}} 
 \delta\left(p^0-\frac{q_2^0}{2}-E_{p-\frac{q_2}{2}}\right)
 \Big\langle{p+\frac{q_2+q_1}{2},p_2-\frac{q_2}{2}\Big|t\Big|
 p+\frac{q_1-q_2}{2},p_2+\frac{q_2}{2}}
 \Big\rangle\Big]_{q_1=q_2=0} \;.\n\\ 
\end{eqnarray} 
Here we consider off-shell contributions, which can be  shown by using  the following equation
\begin{align}
\partial_q^\mu \left. {\frac{1}{2E_{p+\frac{q}{2}}}} \delta\left(p^0+\frac{q^0}{2}-E_{p+\frac{q}{2}}\right) \right|_{q=0}={}& \partial_q^\mu \left. \delta \left(\left(p^0+\frac{q^0}{2}\right)^2-E^2_{p+\frac{q}{2}}\right) \right|_{q=0} \n \\
&{}=p^\mu \delta^\prime (p^2-m^2) \;.
\end{align}
Then all the off-shell contributions to the collision term are collected shown as
\begin{eqnarray}
C_{22,\text{off-shell}}^{(1)}& =& \frac{i\hbar}{2(p^2-m^2)}\, p \cdot \partial \int dP_2 \, 
 f(x,p_1)  f(x,p_2) \n\\
& \times & \bigg\{ 2\int dP_1 dP^\prime\, \frac{1}{i\pi\hbar^2}\left[\Delta_{\text{R}}(p)+\Delta_{\text{R}}^\star(p)\right]
\delta^{(4)}(p+p^\prime-p_1-p_1)  \langle{p,p^\prime|t|p_1, p_2}\rangle\langle{p_1,p_2|t^\dagger|p,p^\prime}\rangle \n\\
&&  + i2\pi\hbar\,  \delta(p^2-m^2) \left[\langle{p,p_2|t^\dagger|p,p_2}\rangle+\langle{p,p_2|t|p,p_2}\rangle\right] \bigg\} \;. \label{c2offshell}
\end{eqnarray}

However, this off-shell contribution  \eq\eqref{c2offshell} exactly cancels 
with the off-shell part on the left-hand side 
of the transport equation \eqref{wigboltz9}. To show it, we  make use of the quasiparticle approximation in 
\eq\eqref{W0}, the left-hand side of \eq\eqref{wigboltz9} then turns into
\begin{equation}
\, p\cdot\partial\, \delta(p^2-m^2-\hbar\delta m^2)f(x,p)
={} \delta(p^2-m^2) p\cdot \partial \, f(x,p) 
+\hbar\frac{1}{p^2-m^2}\, p\cdot \partial \, \mathfrak{M}^{(0)} \;, 
  \end{equation}
where the mass shift at zeroth order reads
\begin{align}
 \mathfrak{M}^{(0)}=
 {}& \frac{i}{2}\int dP_1\, dP_2\, 
 \mathfrak{m}(p_1,p_2,p)\prod_{j=1}^2 f(x,p_j) \;,
    \label{deltaMsq0}
\end{align}
with 
\begin{eqnarray}
 \mathfrak{m}(p_1,p_2,p)&=&
2\int dP^\prime\, \frac{1}{i\pi\hbar^2}
\left[ \Delta_{\text{R}}\left(p\right)+\Delta_{\text{R}}^\star\left(p\right)\right]
\delta^{(4)}(p+p^\prime-p_1-p_1)\langle{p,p^\prime|t|p_1, p_2}\rangle
 \langle{p_1,p_2|t^\dagger|p,p^\prime}\rangle\n \\
& & +i2\pi\hbar p_0\delta(p^2-m^2) \{\delta^{(3)}(\mathbf{p}-\mathbf{p_1})
 [\langle{p_1,p_2|t^\dagger|p,p_2}\rangle  +\langle{p_1,p_2|t|p,p_2}\rangle]+(1\leftrightarrow 2)\}\; ,
\end{eqnarray}
Here we take analogous steps we calculating the local collision term to derive \eq\eqref{deltaMsq0}, as their similar form can be seen from Eqs.~\eqref{deltaM} 
and \eqref{cwww}.

Comparision of  \eq\eqref{c2offshell} with \eq\eqref{deltaMsq0} leads to up to first order
\begin{equation}
C_{22,\text{off-shell}}^{(1)}=\frac{1}{p^2-m^2} \, p\cdot \partial \, \mathfrak{M}^{(0)}\; ,
\end{equation}
which verifies the cancelation of  off-shell parts and indicates
the transport equation involves only on-shell terms for binary collisions up to $\mathcal{O}(\hbar)$.
Thus, we obtain the following kinetic equation for the distribution function $f(x,p)$ for binary collisions
\begin{equation}
 \delta(p^2-m^2)\, p\cdot\partial \, f(x,p)
 = \delta(p^2-m^2)C_{22,\text{on-shell}}[f] \;,
\end{equation}
with 
\begin{equation}
 C_{22,\text{on-shell}}[f]\equiv C_{22,\osl}[f]+\hbar\, C_{22,\os,nl}^{(1)}[f]\;.
\end{equation}
where the term propotional to $\hbar$ comes from the $q_j$ derivatives acting on scattering matrix elements given by 
\begin{eqnarray}
C_{\os,nl}^{(1)}[f]
&=&\left.-2 i\int dP_1\, dP_2\, dP^\prime\,\delta^{(4)}(p+p^\prime-p_1-p_2)\delta(p^2-m^2)\right.\n\\
&&\left.\times f(x,p_2)\partial_{\nu}f(x,p_1)\partial_{q_1}^\nu\left[
\big\langle{{p+\frac{q_1}{2}},{p^\prime}\big|t\big|
	{p_1+\frac{q_1}{2}}, {p_2}}\big\rangle\big\langle{{p_1-\frac{q_1}{2}},{p_2}\big|t^\dagger\big|
	{p-\frac{q_1}{2}},{p^\prime}}\big\rangle\right]\right|_{q_1=0} \n\\
&&+\pi\hbar\int dP_{2} 
\delta(p_1^2-m^2){f(x,p_2)\partial_{\nu} f(x,p_1)}
\partial_{q_1}^\nu \left[\,\big\langle{p+\frac{q_1}{2},p_2\big|
	t\big| p+\frac{q_1}{2},p_2}\big\rangle-\big\langle{p-\frac{q_1}{2},p_2\big|
	t^\dagger\big| p-\frac{q_1}{2},p_2}\big\rangle\right]\Big|_{q_1=0}\;,\n\\
&&+\pi\hbar\int dP_{1} 
\delta(p_2^2-m^2){f(x,p_2)\partial_{\nu} f(x,p_1)}
\partial_{q_1}^\nu\left[  \,\big\langle{p+\frac{q_1}{2},p_1-\frac{q_1}{2}\big|
	t-t^\dagger\big| p-\frac{q_1}{2},p_1+\frac{q_1}{2}}\big\rangle\right]\Big|_{q_1=0}+(1\leftrightarrow 2)\;,\n\\
\label{cmomder1}
   \end{eqnarray}
 where the exchange $1\leftrightarrow 2$ is for the whole expression in front of it.
\end{appendix}

\bibliographystyle{apsrev}
\bibliography{biblio_paper_long}{}

\end{document}